\documentclass[12pt]{iopart}%

\newcommand{\unitx}{\hat{\mathbf{e}}_{\text{x}}}
\newcommand{\unity}{\hat{\mathbf{e}}_{\text{y}}}
\newcommand{\unitz}{\hat{\mathbf{e}}_\text{z}}

\newcommand{\fx}{\hat{F}_{\text{x}}}
\newcommand{\fy}{\hat{F}_{\text{y}}}
\newcommand{\fz}{\hat{F}_\text{z}}
\newcommand{\wrf}{\omega_\text{RF}}
\newcommand{\whf}{\omega_\text{hfs}}
\newcommand{\wMW}{\omega_\text{MW}}

\newcommand{\zRot}{U_\text{z}}
\newcommand{\mFD}{\bar{m}_{F}}
\newcommand{\mF}{m_{F}}

\usepackage{graphics}

\expandafter\let\csname equation*\endcsname\relax
\expandafter\let\csname endequation*\endcsname\relax
\usepackage{amsmath}
\usepackage{subfig}
\usepackage{hyperref}

\usepackage[colorinlistoftodos]{todonotes}

\begin{document}

\title{Bi-chromatic adiabatic shells \\for atom interferometry}

\author{Hector Mas$^{1,2}$, Saurabh Pandey$^{1,3}$, Giorgos Vasilakis$^{1}$ and Wolf von Klitzing$^{1}$}

\address{$^1$ Institute of Electronic Structure and Laser, Foundation for 
Research and Technology-Hellas, Heraklion 70013, Greece}
\address{$^2$ Department of Physics, University of Crete, Heraklion 70013, Greece}
\address{$^3$Department of Materials Science and Technology, University of Crete, Heraklion 70013, Greece}

\vspace{10pt}

\begin{abstract}
We demonstrate bi-chromatic adiabatic magnetic shell traps as a novel tool for matterwave interferometry.
Using two strong  RF fields, we dress the $|1,-1\rangle $ and $ |2,1\rangle$ states of Rubidium Bose-Einstein Condensates thus creating two independently controllable  shell traps.
This allows us to  match the two traps and---using microwave pulses---create a state-dependent clock-type interferometer.
Given the low horizontal confinement of the interferometer, the atoms can be made to spread out thus yielding a 2D sheet, which could be used in a direct imaging interferometer.
This interferometer can be sensitive to spatially varying electric or magnetic fields, which could be DC, AC, RF fields or microwaves.
We demonstrate that the trap-matching afforded by the independent control of the shell traps allows long coherence times which will result in highly sensitive imaging matterwave interferometers.
\end{abstract}

Atom interferometry is a rapidly maturing quantum technology both for fundamental experiments 
and for applications. 
It has been successfully used to measure the Newtonian constant \cite{Rosi2014N}, and to put atom-interferometric constraints on dark energy \cite{Hamilton2015s}.
\emph{Which path} and \emph{delayed choice} experiments have been carried out using atom interferometry \cite{Margalit2015S,Manning2015NP}.
Atom interferometry may be used in tests for sub-gravitational forces on atoms arising from miniature source masses \cite{Jaffe2017NP} and in the search for Ultralight Scalar Field Dark Matter \cite{Geraci2016PRL}.
Simple tests of the weak equivalence principle have been performed using atom interferometry \cite{Zhou2015PRL} and there are proposals for space based extreme accuracy tests at the $10^{-15}$ level \cite{Aguilera2014GAQG} with current projections reaching  $10^{-18}$.
On the applied side, atom interferometers have been used in absolute gravimetry on a ship \cite{Bidel2018NC} and in space \cite{Becker2018N}.
Most precision interferometers still operate in the free-fall-regime, where, e.g.~in the case of acceleration, the precision scales with the square of the interaction time.
As a consequence, the most precise interferometers tend to become very tall, in some cases reaching ten or even one hundred meters in height \cite{Zoest2010S}. %
Even larger interaction times are only possible in zero gravity on parabolic flights or in space \cite{Barrett2016NC}. %
There have been numerous attempts to miniaturize such systems, e.g.~using  shaken lattices \cite{Weidner2018}, partially trapped atoms interferometry with Sr \cite{Zhang2016} and coherent accelerations performed by the Bloch oscillations technique \cite{Andia2013PRA}.
However, fully trapped atom interferometry in a simple magnetic system has remained elusive.
Another aspect is imaging matterwave interferometry, where one uses matterwave interferometry in order to measure the spatial dependence of very minute forces. 
Examples of imaging  interferometers include imaging microwave fields close to the trap \cite{Bohi2010APL} and the cold-atom scanning probe microscope \cite{Gierling2011NN}.

In this paper we demonstrate the basic ingredients for a fully trapped atom interferometer, where a sheet of atoms can be brought close to a surface to image any physical effect to which the trapped states are sensitive to: gravity, dipole forces, electrical fields, RF- and microwaves, etc. 
For this we use adiabatic potentials  to create atomic clouds and BECs in a (quasi) 2D configuration: a strong RF-field dresses atoms in the presence of a  magnetic quadrupole field thus creating a shell-shaped trap, where the atoms are confined to the surface of an axially symmetric oblate spheroid.
Here, we show that by using bi-chromatic RF fields we can create adiabatic shell traps for both $F=1,m_F=-1$ and $F=2,m_F=+1$ and manipulate them independently.
Here, $F$ is the quantum number of the hyperfine state and  $m_F1$ its magnetic hyperfine state.

This allows us to create two identical shell traps for the two states and then to couple them using microwave photons.
This provides the ideal starting point for a 2D imaging interferometer:
The low trapping frequency in the horizontal directions allows one to create an extended two-dimensional potential, which can act as an imaging sensor for minute fields and forces.
We demonstrate that despite the difficulties of matching the two traps one can achieve long coherence times even in the presence of large background fluctuations, e.g.~in the DC magnetic field, thus demonstrating the viability of a shell-trap imaging interferometer.

We begin by giving a background to the theory of the simple RF-dressed shell potentials in Sec.\,\ref{sec:ShellTrap} and its bi-chromatic extension Sec.\,\ref{sec:BiChromaticShellTrap}. 
We  address the question of microwave based beam splitters in Sec.\,\ref{sec:BeamSplitters} and discuss the sources of dephasing and decoherence in Sec.\,\ref{sec:DephasingSources}. Finally in the conclusion (Sec\,\ref{sec:ConclusionsAndOutlook}) we also present some future applications.

\section{The RF-dressed shell trap}
\label{sec:ShellTrap}

We consider the case of a $^{87}$Rb atom in the presence of a static quadrupole field 
$\boldsymbol{B}_\text{q}\left(\mathbf{r}\right)=\alpha (x\unitx+y \unity-2 z\unitz)$, where $\alpha$ is a magnetic quadrupole gradient and $(\unitx,\unity, \unitz)$ are the unit vectors. A homogeneous RF-field of arbitrary polarization $\boldsymbol{B}_{\textrm{RF}}(t)$ and frequency $\wrf$ is also applied. The Hamiltonian for weak fields $|g_\text{F} \mu_\text{B} \boldsymbol{B}_\text{q}(\mathbf{r})|\ll \hbar \omega_\text{hfs}$ (where $\omega_\text{hfs}$ is the hyperfine splitting frequency, and $\mu_{B}$ is the Bohr magneton) reads:
\begin{equation}
H = \frac{A_{\text{hfs}}}{\hbar^{2}} \boldsymbol{I}\cdot\boldsymbol{J} + \frac{\mu_\text{B}}{\hbar}(g_I \boldsymbol{I} + g_J \boldsymbol{J}) \cdot (\boldsymbol{B}_\text{q}(\mathbf{r})+\boldsymbol{B}_{\textrm{RF}}(t)),
\label{eq:H_uncoupled}
\end{equation}

\noindent where $A_{\text{hfs}}$ is the hyperfine constant of $^{87}$Rb, $\boldsymbol{I}$ and $\boldsymbol{J}$ are the nuclear and electronic angular momentum operators, with $g_{I}$ and $g_{J}$ being their respective gyromagnetic ratios. For weak magnetic fields, the total angular momentum  $\boldsymbol{F}=\boldsymbol{I}+\boldsymbol{J}$ with eigenstates in a coupled basis $|F,\mF\rangle$ can be used instead of $|I,m_{I},J,m_{J}\rangle$. For $^{87}$Rb at the electronic ground state ($I=3/2$, $J=1/2$) there are two hyperfine subspaces with $F=I\pm 1/2$, and hyperfine Zeeman sub-states given by $\mF=-F\dots 0 \dots F$ in each one. The total dimension is $(2J+1)(2I+1)=8$. The total angular momentum gyromagnetic factor $g_{F}$ is (see \cite{Steck2001} and references therein)

\begin{equation}
g_{F}=g_{J}\frac{F(F+1)-I(I+1)+J(J+1)}{2F (F+1)}+g_{I}\frac{F(F+1)+I(I+1)-J(J(J+1))}{2F(F+1)}
\label{eq:gFdefinition}
\end{equation}

\noindent with $g_{J}=2.002331$ and $g_{I}=-0.000995$. Eq.~\ref{eq:gFdefinition} results in different $g_{F}$ for $F=1,2$ ($g_{1}=-0.5018$ and $g_{2}=0.4998$) \cite{Steck2001}.
Eq.~\ref{eq:H_uncoupled} is generally solved and simplified by applying the Rotating Wave Approximation (RWA) and by using a \textit{dressed} basis $|F,\mFD\rangle$, of the same dimension as $|F,\mF\rangle$. The result, including gravity, is the adiabatic dressed potential~\cite{Zobay2001}:

\begin{equation}
V^{F}\left(\mathbf{r}\right)=s\left(I+\frac{1}{2}\right)\frac{\hbar \omega_\text{hfs}}{2}+ s\mFD \hbar  \sqrt{\delta_{F}^{2} + \Omega_{F}^{2}\left(\mathbf{r}\right)}+M g z
\label{eq:dressedPotentiala}
\end{equation}

\noindent where $s=g_{F}/|g_{F}|$ is the sign of $g_{F}$, with 
  $s=-1$ for $F=1$ and  $s=+1$ for $F=2$.  
  $M$ is the atomic mass of $^{87}$Rb and $g$ is the earth's gravitational acceleration.
The detuning of the RF from the resonance is $\delta_{F}=\Omega_\text{L}^{F}\left(\mathbf{r}\right)-\omega_{F}$ where $|\hbar \Omega_\text{L}^{F}(\mathbf{r})|= |g_{F} \mu_{\text{B}} \boldsymbol{B}_{q}(\mathbf{r})|$ is the Larmor frequency and $\omega_{F}$ is the RF-frequency $\wrf$ that dresses each manifold, i.e. \,$\omega_{1}$ for F=1 and $\omega_{2}$ for F=2. The Rabi coupling, $\Omega_{F}(\mathbf{r})$, varies with the position and depends on the spin gyromagnetic ratio \cite{Garraway2015}. Please note, that the sign of $g_{F}$ determines which local polarization component of the RF couples the $m_F$ states, i.e.~for the absorption of one rf-photon $(\Delta m_F=+1)$ one requires $\sigma_{-}$ for the $F=1$ and $\sigma_{+}$ for the $F=2$ state. 
Since the $F=1$ and $F=2$ manifolds are interacting with mutually orthogonal RF-polarizations, they can be addressed entirely independently. A circularly polarized RF-field in the laboratory frame can be written as:

\begin{equation}
\boldsymbol{B}_\text{RF}=B_{\text{RF}}\left(\cos{\left(\omega_{F} t\right)}\unitx+s\sin{\left(\omega_{F} t\right)}\unity\right) 
\label{eq:monochrombfield}
\end{equation}

\noindent Each sense of rotation, specified with $s=\pm 1$ and defined with respect to the symmetry axis of the quadrupole field, couples atoms to either of the F manifolds. The local coupling with respect to the spatially dependent quantization axis defined by the direction of $\boldsymbol{B}_{q}(\mathbf{r})$ is:

\begin{equation}
|\Omega_{F}\left(\mathbf{r}\right)|=\frac{\Omega_{\text{RF}}}{2}\left(1-\frac{2z}{\sqrt{\rho^{2}+4z^{2}}}\right)
\label{rabicoupling}
\end{equation}

\noindent where $\hbar \Omega_{\text{RF}}=|g_{F}|\mu_{\text{B}}B_{\text{RF}}$ is the maximum Rabi frequency (note that all axes have an origin at the zero of the quadrupole field). The resulting potential is an iso-magnetic surface defined by the resonance of $\omega_{F}$ and $\boldsymbol{B}_{q}\left(\mathbf{r}\right)$ \cite{Lesanovsky2006}. Trapping requires low-field-seekers, i.e.~states with $g_{F} \mF> 0$, when $\gamma=\frac{M g}{2 \mF g_{F} \mu_{\text{B}} \alpha}<1$ \cite{Merloti2013}.
The trapping frequencies close to the trap center are:
\begin{eqnarray}
\omega_{\rho}= \sqrt{\frac{g}{4 z_{0}}}\left(1-\frac{|\mF|\hbar \Omega_{\text{RF}}}{2 m g z_{0}}\sqrt{1-\gamma^{2}}\right)^{\frac{1}{2}} \label{eq:shelltrapfreqrad} \\
\omega_\text{z}=\frac{2 |g_{F}| \mu_{B} \alpha}{\hbar} \sqrt{\frac{|\mF| \hbar}{M \Omega_{\text{RF}}}}\left(1-\gamma^{2}\right)^{\frac{3}{4}}
\label{eq:shelltrapfreqaxial}
\end{eqnarray}

\noindent where:

\begin{equation}
z_{0}=\frac{\hbar \omega}{2\alpha \mu_{\text{B}} |g_{F}|}\left(1+\frac{\gamma}{\sqrt{1-\gamma^{2}}}\frac{\Omega_{\text{RF}}}{\omega}\right)
\label{eq:trapposition}
\end{equation}
is the vertical position of the trap. The values of $\gamma$ for the the spin states $|1,\mFD=-1\rangle$ and $|2,\mFD=1\rangle$ states differ only by a small amount, e.g. $\gamma_{1}/\gamma_{2}=g_{2}/g_{1}\approx0.996$.
This small difference causes the two traps to have both different positions $z_0$  and trapping frequencies. 
Therefore, any superposition between the two states  will  dephase rapidly, thus  severely limiting any interferometric measurement using these two states.

\section{The Bi-chromatic Shell Trap \label{sec:BiChromaticShellTrap}}

As stated above, the two manifolds $F=1$ and  $F=2$ are dressed by  RF-fields, which are mutually orthogonal and can have different frequencies and/or amplitudes.
The  mismatch of the traps can be overcome at least partially using \emph{bi-chromatic} adiabatic  potentials. 
The RF-dressing fields used in the bi-chromatic shell trap are composed of one  $\sigma_{-}$ polarized field with $(\omega_{1}, \Omega_{1})$ and one $\sigma_{+}$ polarized field with $(\omega_{2},\Omega_{2})$. They can be generated from  rf-coils placed in the x- and y-direction with the linearly polarized fields as:

\begin{equation}
\begin{split}
\boldsymbol{B}_\text{\text{RF}}(t)=
\boldsymbol{B}_{1}(t)+\boldsymbol{B}_{2}(t)=
B_{\text{x1}} \cos{\left(\omega_{1}t\right)}\unitx-
B_{\text{y1}} \sin{\left(\omega_{1}t\right)}\unity
\\+
B_{x2} \cos{\left(\omega_{2}t\right)}\unitx +
B_{y2} \sin{\left(\omega_{2}t\right)}\unity
 \end{split}
 \label{eq:bifield}
 \end{equation}

 \noindent which results in two circularly polarised components at the position of the atoms when the amplitude of both frequency components at the RF-coils is the same.

The traps for the two states have independently tunable position and trapping frequencies. 
Perfect matching would be achieved when the positions and both trapping frequencies are identical for both states at the same time. 
Unfortunately, this cannot be achieved simultaneously  for both states, see Eqs.(\ref{eq:shelltrapfreqrad},\ref{eq:shelltrapfreqaxial},\ref{eq:trapposition}). 
It is possible, however, to  perfectly match $z_{0}$ and $\omega_{z}$ and at the same time keep the difference in $\omega_{\rho}$ 
very small. 
For a given $\alpha$ and by fixing the parameters for one trap ($\omega_1$ and $\Omega_1$), we can calculate  the ones for the other:

\begin{eqnarray}
\Omega_{2}=\left(\frac{g_{2}}{g_{1}}\right)^{2}\left(\frac{1-\gamma_2^2}{1-\gamma_1^2}\right)^{3/2}\Omega_1
\label{eq:matchingcondrabi}\\
\omega_{2}=\frac{g_{2}}{g_{1}}\left(\frac{\gamma_{1} \Omega_{1}}{\sqrt{1-\gamma_{1}^2}}+\omega_{1}\right)-\frac{\gamma_2 \Omega_2}{\sqrt{1-\gamma_{2}^2}}\label{eq:matchingcondfreq}
\end{eqnarray}

\begin{figure}
\begin{center}
\resizebox{0.8\textwidth}{!}{%
  \includegraphics{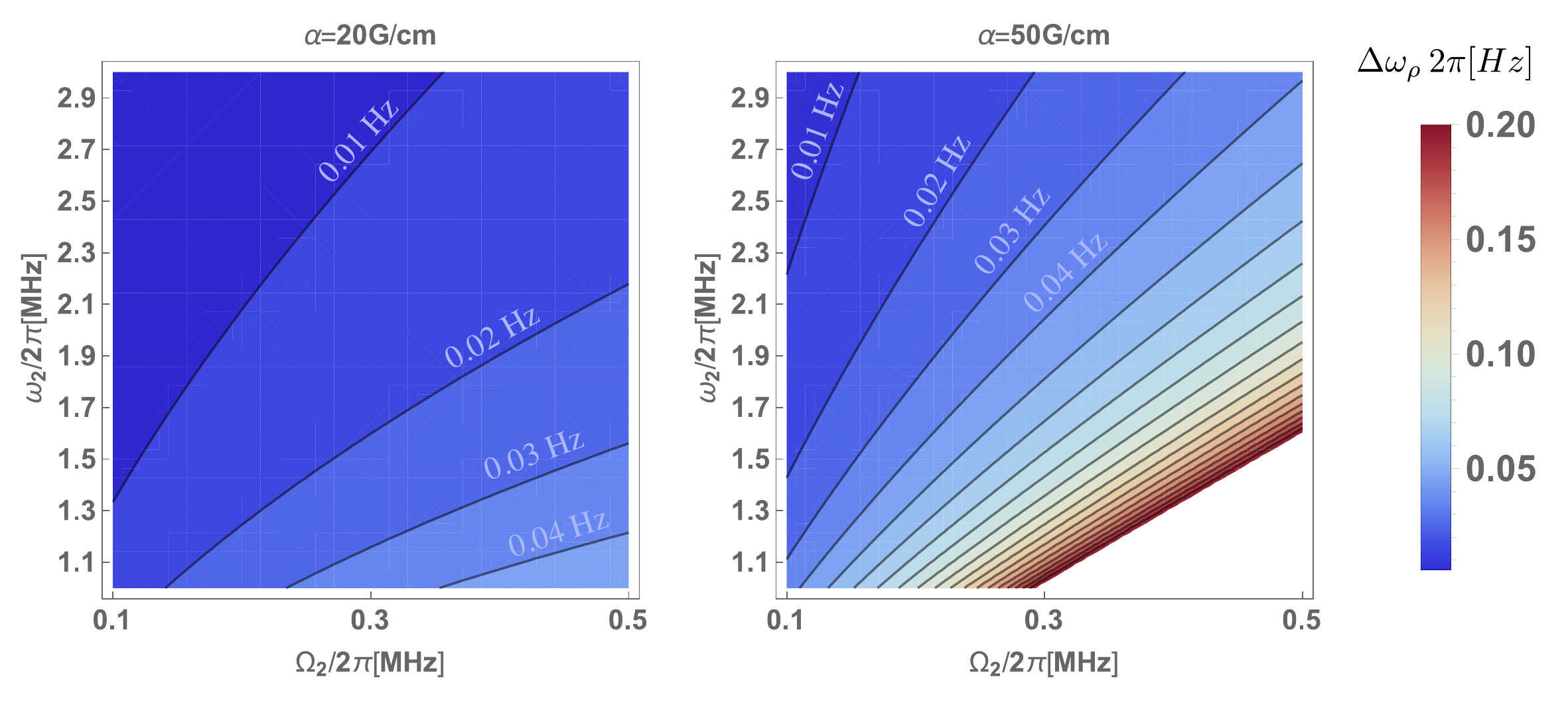}}
  \caption{Difference in radial frequency between the two traps, calculated as $\Delta \omega_{\rho}/2\pi$. For any $B_q$, the position $z_0$ and the axial trapping frequency $\omega_{z}$ are perfectly matched. One can match $\omega_{r}$ but only for asymptotically non-physical parameters, e.g. for very large RF-frequencies and for very small Rabi frequencies and quadrupole gradients.}
  \label{fig:radialfreqdiff}
  \end{center}
\end{figure}

\noindent The full analytic expression for the resulting radial frequency difference is too complex to be printed here. 
Fig.~\ref{fig:radialfreqdiff} shows the difference in radial frequencies $\Delta \omega_{\rho}/2\pi$  for a range of Rabi and RF frequencies assuming the matching conditions of Eq.(\ref{eq:matchingcondrabi}). 
It is clear that the frequency difference decreases with  increasing radio frequency $(\omega_{\textrm{RF}})$ and decreasing gradient $(\alpha)$ and Rabi frequency $(\Omega_{\textrm{RF}})$.  
We find, for example, that $\Delta \omega_{\rho}/2\pi$ can be as low as $10$\,mHz for Rabi frequencies of the order of $100$\,kHz, RF-frequencies of the order of 2.5MHz and gradients of the order of $20$\,Gpcm. 
Assuming only a single vibrational level is occupied, this implies a de-phasing time of the order of 10\,s.

\section{Microwave beam splitters for adiabatic potentials \label{sec:BeamSplitters}}
%

\iffalse
BS
Fluctuations
 DC
-- Homogeneous
-- QP
AC
-- RF
-- mw
\fi

In the atom interferometer based on the adiabatic traps \cite{Navez2016}, beam splitters are microwave transitions between the clock states.
This adds a term proportional to $\boldsymbol{B}_\text{MW}(t)$ to the total magnetic field term in Eq.~\ref{eq:H_uncoupled}.
The solution of this Hamiltonian for a weak MW-field that acts as a probe on the RF-dressed energy levels leads to inter-manifold transitions \cite{Haroche1970,Nottingham,Luksch2018} on a spectrum of 7 groups (spaced by the RF-dressing frequency) of 5 transitions (spaced by the RF-dressing Rabi frequency), when the initial state is $|F=1,\bar{m}_{F}=-1\rangle$.
These are given by the condition: 
$
	\omega_{\text{MW}} = \omega_\text{hfs} + k\,\Omega_{\text{RF}}-n\,\omega_\text{RF} ,
	\label{eq:resonanceCondition}
$
with $n=-3...0...3$ and $k=-3...0...1$.
For all $n$ and $k=0$ transitions between $|1,-1\rangle \rightarrow |2,1\rangle$ occur. Note that we have written this simplified formula for the case where both manifolds are dressed by Rabi frequencies of equal magnitude (see \cite{Nottingham} for a detailed study). 
Concretely, the transition with $n=0,k=0$ is coincident with the hyperfine splitting \cite{Nottingham}. 
In practice, this transition is shifted both by the non-linear dependence of the eigen-energies on the magnetic field and on the difference in magnitude of $g_1$ and $g_2$. 
In addition, we find that the cross-coupling between $B_{2}$ ($B_{1}$) and $F=1$ ($F=2$) induced by the polarization defects results in a periodic modulation of the energy splitting that causes multi-photon transitions \cite{Silveri2017, Oliver2005}, and also time-averaged reshaping of the traps.
In the ideal case, the bi-chromatic shell should exhibit the spectral response explained above, i.e.\,7 peaks close to $\omega_\text{MW}=\omega_\text{hfs}+n\omega_\text{RF}$.
However, we observe a finite number of harmonics separated by $\Delta \omega_{F} =|\omega_{1}-\omega_{2}|$.
We explain these multi-photon transitions with a semi-classical two-level toy model in \ref{app:multiphoton}.

\begin{figure*}
          \begin{minipage}{.47\textwidth}\centering
     \subfloat[][]{\includegraphics[height=1.9in]{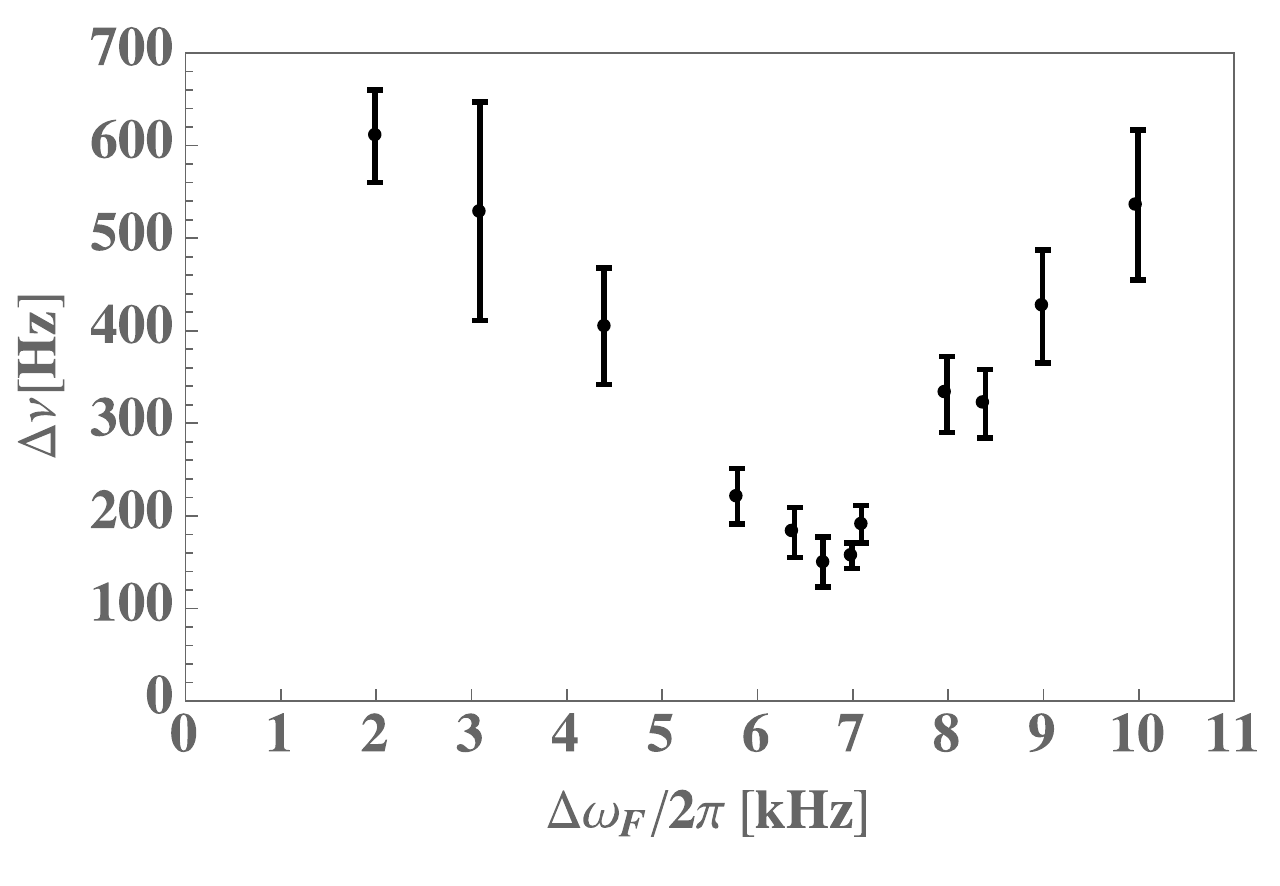}\label{fig:tuningfrequencya}}
     \end{minipage}
          \begin{minipage}{.47\textwidth}\centering
     \subfloat[][]{\includegraphics[height=1.9in]{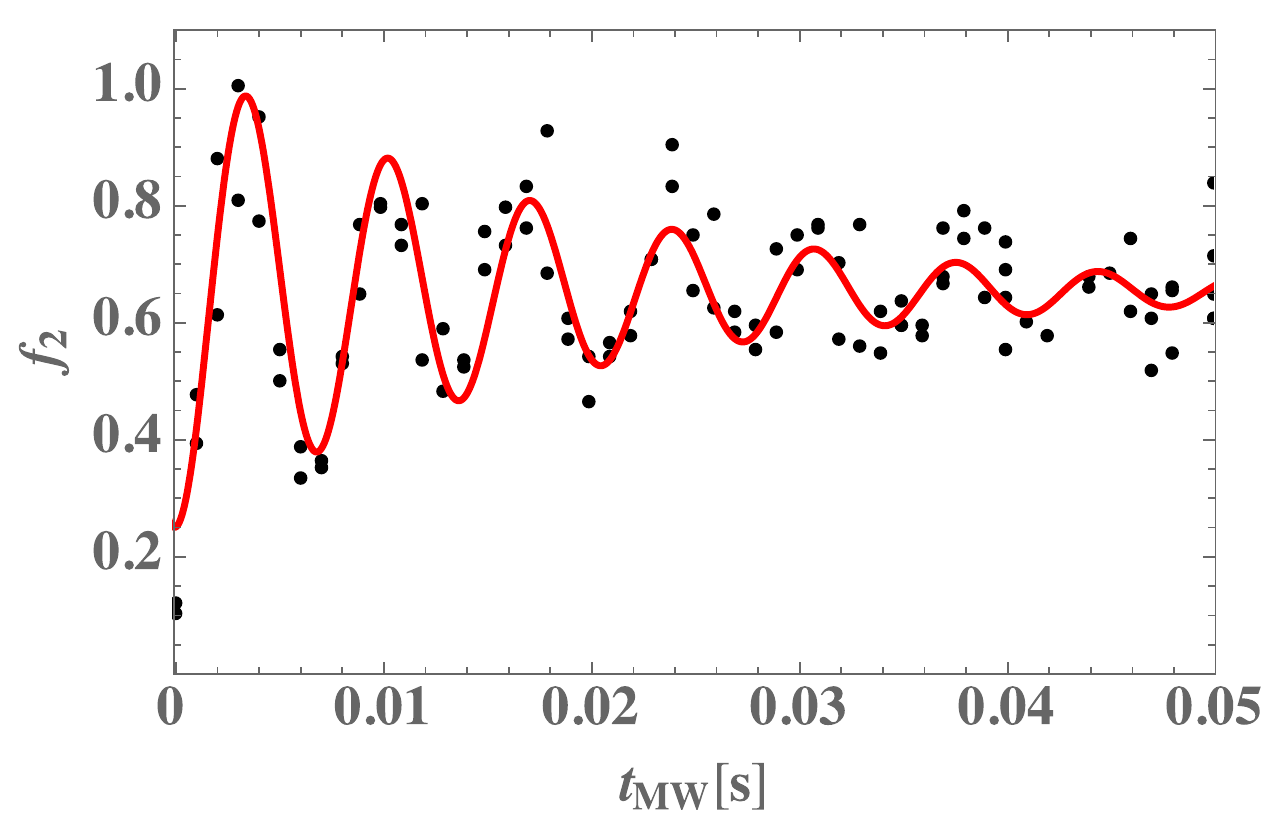}\label{fig:tuningfrequencyb}}
     \end{minipage}%
     \caption{(a): measured line-width $\Delta \nu$ in the $n=1$ transition with a microwave pulse of $10$\,ms for the clock transition in the dressed states for different values of $\Delta \Omega$. $\Omega_{1}/2\pi\approx\Omega_{2}/2\pi\approx  250$\,kHz, $\alpha=45$\,G/cm. (b): Example of Rabi oscillations in the bi-chromatic adiabatic potential.}
\label{fig:tuningfrequency}
\end{figure*}

In Section~\ref{sec:BiChromaticShellTrap} we found an approximation for the $\Delta \omega_F$ that matches the shell traps of the two states.
In practice, however, the exact matching is affected by non-linear Zeeman shifts and the experimental  imperfections in the generation of the (RF-)field.
In this section, we present a method to match the two traps experimentally  by spectroscopic means: we choose a pair of frequencies $\omega_{1}$, $\omega_{2}$ close to the prediction of Eqs.~(\ref{eq:matchingcondrabi},\ref{eq:matchingcondfreq}) for $\alpha=45$\,G/cm and $\Omega_{1}/2\pi\approx \Omega_{2}/2\pi \approx 250$\,kHz (see Appendix~\ref{app:traploading} for an explanation of the sample preparation and trap loading).
We then find the longest coherence times by measuring the line-width of the transition  for different values of $\Delta \omega_\text{F}$. To do this, we load the bi-chromatic shell trap for several pairs of $\omega_{1}$ and $\omega_{2}$ and then we fit Lorentzian curves to the number of atoms transferred to the upper state as we scan the microwave frequency $\omega_{\text{MW}}/2 \pi$. By finding the combination of $\omega_{1}$ and $\omega_{2}$ that gives the minimum linewidth, we can thus match the two traps (see Fig.~\ref{fig:tuningfrequency} a)), at which point  we observe a reduction in line-width by of the order of $\times 10$ with respect to the mono-chromatic shell (measured elsewhere \cite{Nottingham}).

In the mono-chromatic shell trap we observe very fast decay curves and no Rabi oscillations, and a line width of $\Delta \nu\approx 1000$\,Hz. 
In contrast to this, in the bi-chromatic shell we measured a line-width of $\Delta \nu \approx100$\,Hz for $\Delta t_\text{MW}=10$\,ms, and we observe Rabi oscillations with a decay time of $>10$\,ms. Fig.~\ref{fig:tuningfrequency} b) shows an example of the Rabi oscillations in the bi-chromatic shell trap, with $\alpha =50$\,G/cm, $\omega_{1}/2 \pi=2.294336$\,MHz, $\omega_{2}/2 \pi=2.285238$\,MHz and $\Omega_{1}/2\pi\approx \Omega_{2}/2\pi\approx 270$\,kHz. In the figure, $f_{2}$ is the normalized atom number in the dressed $|2,1\rangle$ state. We fit the data to the the following function $f_{2}=f_{o,2}+(f_{s,2}/2)\left(1- e^{-t/\tau}   \cos{\left(\Omega_{\textrm{MW}} t_{\textrm{mw}}\right)}\right)$. In this equation, $\Omega_{\textrm{MW}}$ is the MW driving field Rabi frequency in the RF-dressed TLS, and $\tau$ is the decay time of the Rabi oscillations amplitude for this transition.
A fit to the data in Fig.~\ref{fig:tuningfrequency} b) yields $\Omega_{\textrm{MW}}/2\pi=146\pm 1$\,Hz and $\tau=18\pm 2$\,ms. The term $f_{0}$ is an offset that accounts for atoms that are previously transferred to the $|2,1\rangle$ due to an imperfect switching on of the MW radiation. The fields at $\omega_{1}$ and $\omega_{2}$ are produced from two separate two-channel sources and then mixed before the amplifiers that are connected to the RF-antennae in the x- and y-direction (see~\ref{app:rfgeneration}).

\section{Dephasing and decoherence\label{sec:DephasingSources}}
\begin{figure}
 \includegraphics[width=1 \linewidth]{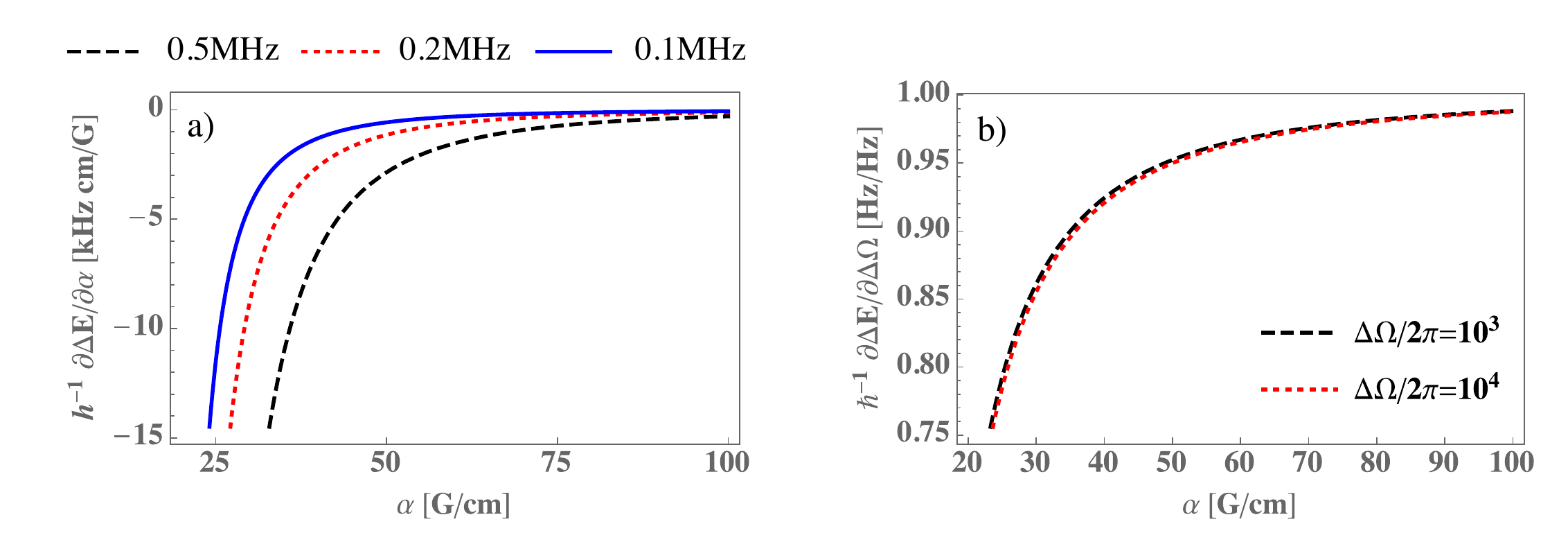}
\caption{ (a) We show the sensitivity of the transition to fluctuations in the gradient for various gradients and various $\Omega_{2}$ for fully matched traps (i.e.\,Eqs.~(\ref{eq:matchingcondrabi},\ref{eq:matchingcondfreq}) are fulfilled). (b) We show the susceptibility of the transition associated to fluctuations in the difference in Rabi frequency, that is $\Delta \Omega= \Omega_{2}-\Omega_{1}$.
}
\label{fig:noisesourcesfig}
\end{figure}

The RF frequencies are generated with reference to an atomic clock and as such do not contribute to the dephasing. 
The energy levels of the atoms in the shell trap are insensitive to changes in the homogeneous DC-fields since their only effect is to move the center of the quadrupole trap.
However, they are  sensitive to fluctuations in the gradient of the quadrupole field and in the Rabi frequencies of the RF-fields, especially to the difference of the Rabi frequencies of the two states, i.e.\,$\Delta \Omega= \Omega_{2}-\Omega_{1}$.
The transition is relatively insensitive to fluctuations in the Rabi frequency as long as $\Delta \Omega$ remains stable.
Fig.~\ref{fig:noisesourcesfig} a) shows the sensitivity of the transitions with respect to fluctuations of the magnetic gradient field.
We write them as $\partial \Delta E/\partial \alpha$, where $\Delta E=V^{2}-V^{1}$ is the difference in energy between the two adiabatic potentials retrieved from Eq.~\ref{eq:dressedPotentiala}, assuming the traps are matched in position and axial trapping frequency at all times following Eqs.~(\ref{eq:matchingcondfreq},\ref{eq:matchingcondrabi}).
Clearly, this sensitivity increases with decreasing gradients.
Fig.~\ref{fig:noisesourcesfig} b) shows the sensitivity of the transition on fluctuations in the difference of Rabi frequencies $\Delta \Omega$, again assuming that the traps are perfectly matched and that the fluctuations in the two RF-fields are uncorrelated.
In this case, the sensitivity is smaller for lower quadrupole gradients $\alpha$.

\subsection{Sensitivity to homogeneous fields
\label{sec:LineShiftWithBiasField}}

\begin{figure}
\begin{center}
\resizebox{0.5\textwidth}{!}{%
  \includegraphics{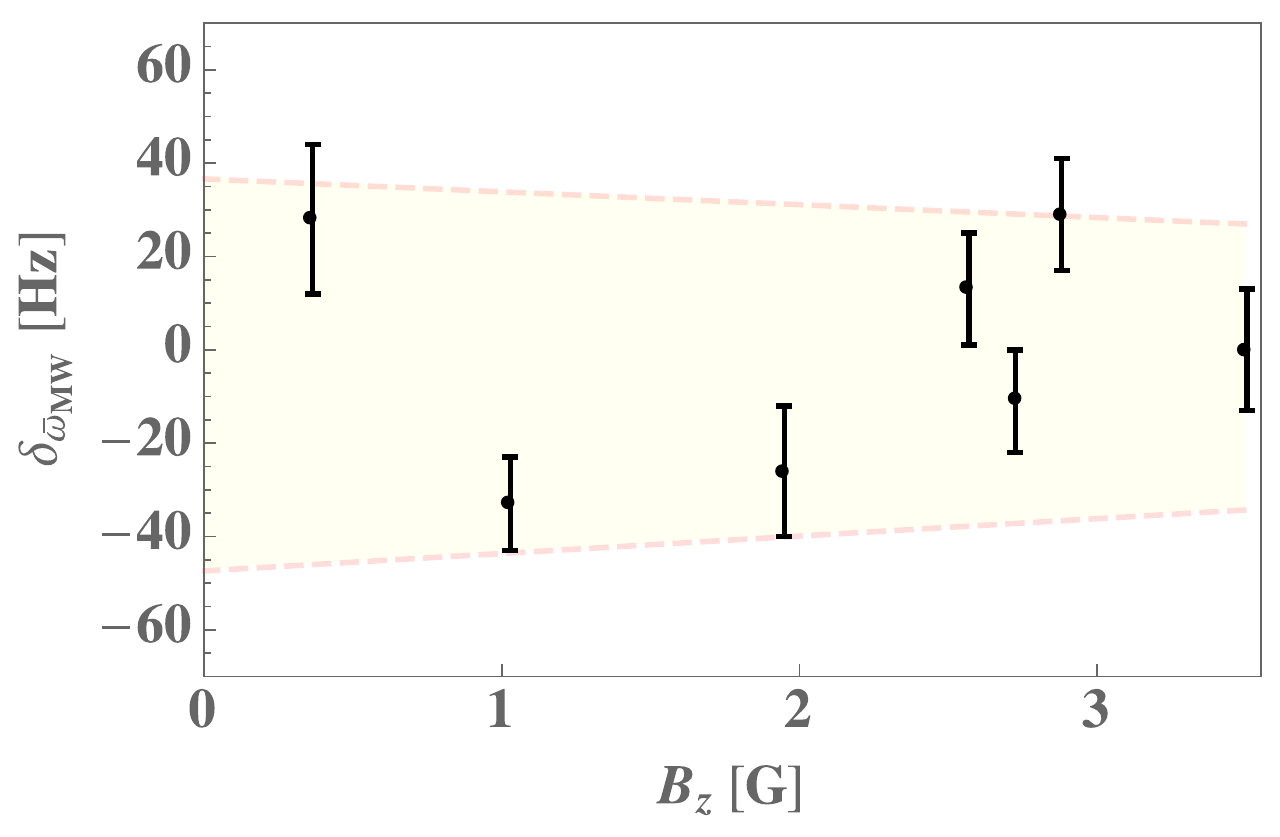}}
\caption{Measured $\delta_{\bar{\omega}_\text{MW}}=(\omega_{\text{MW}}-\bar{\omega}_{\text{MW}})/2 \pi$ where $\bar{\omega}_\text{MW}/2\pi=\omega_\text{hfs}+2\pi\, 6210$\,Hz
is the mean value of the measured transition frequencies.
The trap parameters were $\omega_{1}/2\pi=2.294336$\,MHz, $\omega_{2}/2\pi=2.285238$\,MHz, $\Omega_{1}\approx \Omega_{2}\approx240$\,kHz, $\alpha=70$\,G/cm. The MW pulse lasted $\Delta t_{\text{MW}}=2$\,ms, much less than one Rabi cycle.
We fit a linear slope to this data set and find it to be $(2\pm10)$\,Hz/G, with an offset of $(-4\pm24)$\,Hz. The yellow area represents the confidence interval given by the statistical uncertainty of the fit.
}
\label{fig:bfieldshift}
\end{center}
\end{figure}

Compared to other magnetically trapped atom-clock interferometers, the shell trap has the considerable advantage that homogeneous DC magnetic field fluctuations do not influence the transition, thus eliminating the need for DC magnetic shielding.
The reason behind this is that the shell trap is based on a magnetic quadrupole field. 
Any homogeneous field simply shifts the trap center without affecting its trapping parameters.
In order to check that this shift does not affect the trap, e.g.~due to field inhomogeneities, we measured the clock frequencies of the dressed  $|1,-1\rangle\rightarrow|2,1\rangle$ transition (by finding the center of Lorentzian fits to the atomic population in the dressed $|2,1\rangle$ after short pulses) for extreme variations in homogeneous background field ($0 \leq B_{\text{z}} \leq 3.5$\,G).
For this we loaded a bi-chromatic shell with $\omega_{1}/2\pi=2.294336$\,MHz, $\omega_{2}/2\pi=2.285238$\,MHz, $\Omega_{1}/2\pi\approx \Omega_{2}/2\pi\approx 240$\,kHz and $\alpha=70$\,G/cm (see~\ref{app:traploading}).
Then we drove the transition at $n=0, k=0$ ($\Delta t_{\text{MW}}=2$\,ms) for various field strengths $B_{\text{z}}$.
Fig.~\ref{fig:bfieldshift} shows the detuning $\delta_{\bar{\omega}_\text{MW}}=(\omega_{\text{MW}}-\bar{\omega}_{\text{MW}})/2 \pi$, with the shaded yellow area corresponding to the confidence interval given by the statistical uncertainty of a linear fit to the data points. Here $\bar{\omega}_{\text{MW}}$ is the mean value of all measurements of the transition frequency $\omega_{\text{MW}}/2\pi$ for different homogeneous fields $B_{z}$.
This agrees with the experimental observation that there is no shift in the transition frequency of $|1,\mFD=-1\rangle \rightarrow |2,\mFD=1\rangle$ in the range from zero to 3.5\,G. When we fit a linear slope to the set of data in Fig.~\ref{fig:bfieldshift}, we find a near zero slope ($2\pm10$\,Hz/G) and zero offset ($4\pm24$\,Hz).

\begin{figure}
\begin{center}
\resizebox{0.5\textwidth}{!}{%
  \includegraphics{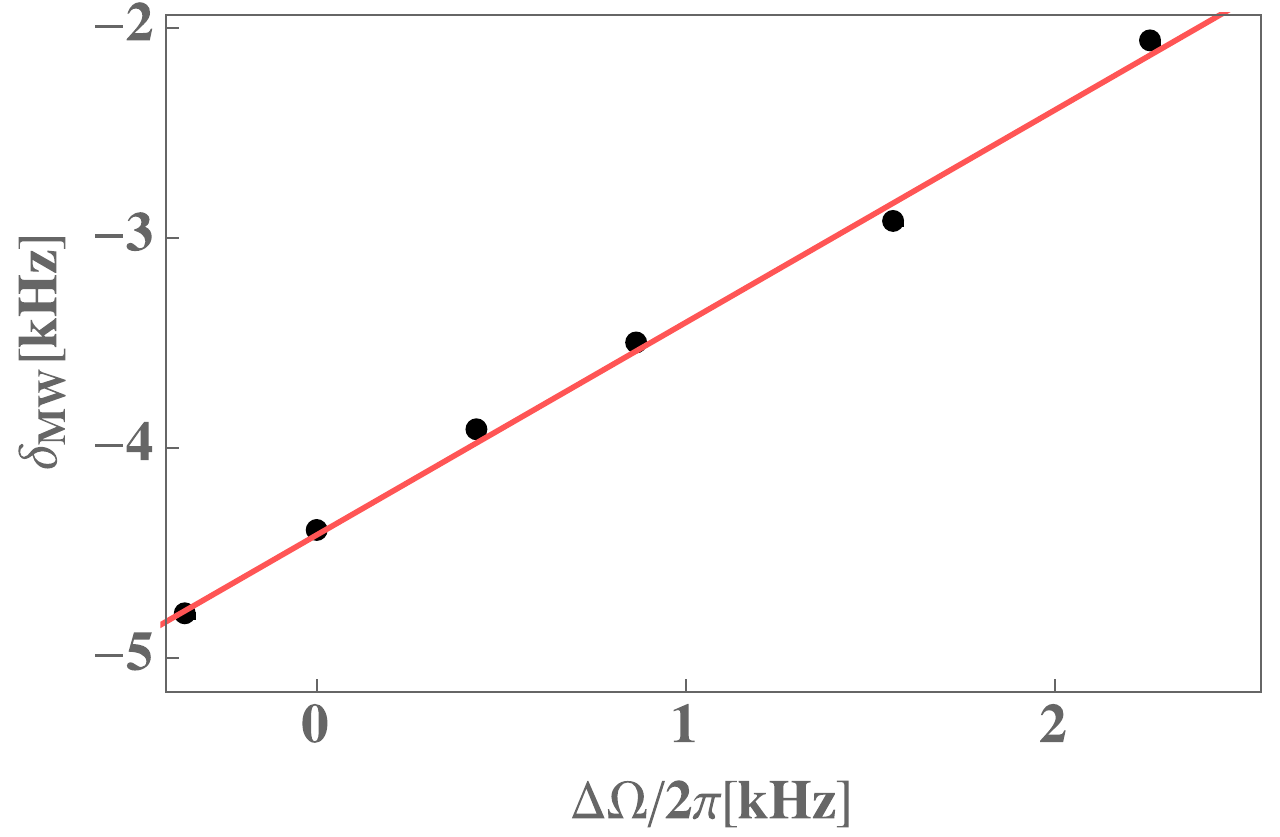}}
\caption{Measured shift for the dressed $|1,\mFD=-1\rangle \rightarrow |2,\mFD=1\rangle$ transition for several values of the difference in Rabi frequency $\Delta \Omega=\Omega_{2}-\Omega_{1}$ between the $V^{1}$ and $V^{2}$ shell traps.
The solid red line is a fit that yields a slope of $1.01\pm 0.03$\,Hz/Hz with an offset of $-4412\pm 20$\,Hz. 
The statistical error in the frequencies is much smaller than the dot-size.
There is a systematic error of about $ 1$\,kHz due to the uncertainty of the determination of $\Delta \Omega/2\pi$ .
}
\label{fig:DeltaOmegaShift}
\end{center}
\end{figure}

\subsection{Line shift with $\Delta \Omega$}
\label{sec:shiftRabi}

We can estimate from Eq.\,\ref{eq:dressedPotentiala} the line shift with the difference in Rabi frequency ($\Delta \Omega$) under the assumption that the position of the two traps is matched.
This shift becomes directly proportional to $\Delta \Omega$ when the gradient is high (see Fig.~\ref{fig:noisesourcesfig} b)).
Fig.\,\ref{fig:DeltaOmegaShift} shows the transition frequency $\delta_\text{MW}=(\omega_\text{MW}-\whf)/2 \pi$ for several $\Delta \Omega$.
The Rabi frequencies had been previously calibrated by measuring the transitions induced by an additional weak RF-field \cite{GarridoAlzar2006}. To do this, we loaded a one-frequency shell trap at different RF-amplitudes and measured atomic transitions to untrapped states. Atomic transitions to untrapped states in this case occur at $\Omega_{\text{RF}}$ and $\wrf\pm \Omega_{\text{RF}}$. From here one can calibrate the RF-field amplitude.  
We also determined $\Omega_1$ and $\Omega_2$ from the frequency difference between  transitions involving different number of RF photons (see Eq.\,\ref{eq:resonanceCondition}).
Initially, we set $\Omega_{1}/2\pi\approx \Omega_{2}/2\pi \approx 370$\,kHz and $\alpha=22$\,G/cm
calculated from Eqs.~(\ref{eq:matchingcondrabi},\ref{eq:matchingcondfreq}), and subsequently fine-tuned the frequency difference to compensate for the non-linear effects and imperfections in the RF-generation.
The final frequencies were $\omega_{1}/2 \pi=2.267900$\,MHz and $\omega_{2}/2 \pi=2.258910$\,MHz.
The red line is a linear fit.
Within the accuracy of the fit, the shift in angular transition frequency is equal to the shift in $\Delta \Omega$, which implies a sensitivity to the different amplitude of the driving field of 700\,kHz/G. Note that, from Fig.~\ref{fig:noisesourcesfig} b) we would expect a slightly smaller line-shift at a gradient of $\alpha=22$\,G/cm. However, in this experiment it is not clear that the traps are matched to the degree of precision that the calculation in Fig.~\ref{fig:DeltaOmegaShift} assumes. Moreover, we have not taken into account non-linear terms, which may not be negligible. 
In any case, this poses strict limits on the stability of the RF generation.
For example, a coherence time of 1\,s  would require a stability of the RF magnetic field that defines $\Delta \Omega$ of, at least, $\sim 10^{-5}$.
Drifts in the gain of amplifiers or in the quality factor of resonators make such stability very hard to achieve in the case of linear polarization.
In the case of the shell traps, though,  this condition applies to the two circularly polarized fields of orthogonal polarization. 
Both are generated from the the same two pairs of Helmholtz coils and differ only slightly in frequency, with the polarization determined by the relative phase of the RF in the two coils. 
A change in any single coil or amplifier will cause a change in the polarization and amplitude of both RF components equally, thus considerably reducing its impact on the transition frequency.
\section{Conclusions and Outlook 
\label{sec:ConclusionsAndOutlook}}

We introduced bi-chromatic adiabatic potentials as a novel system for guided/trapped atom interferometry. 
We introduced a simple model for the matching condition for the two shells and demonstrated a method of fine-tuning, which results in approximately an order of magnitude reduction in the line-width of the transition between the dressed states $|1,-1\rangle$ and $|2,1\rangle$ of $^{87}$Rb compared to the single frequency case.
The resulting coherence of time of 20\,ms has been achieved for a magnetically sensitive state, without any shielding or active stabilization of either DC or RF-fields. 
First calculations indicate that our set-up should be adapted for lower RF-frequencies and higher RF-amplitudes, both within reach experimentally.
We expect that with proper care the current coherence time can be extended by one or two orders of magnitude.
The demonstrated insensitivity of the transition to DC-fields at least up to 3\,G allows one to move the shell simply by applying a homogeneous field.
Applications will include an imaging atom-interferometric surface probe for DC and AC fields.
For this, one would load a BEC into a large-diameter shell-trap, approach it to a surface, and image the spatial dependence of the atom-interferometric sequence. 
Further theoretical and experimental work will be required  for the shell trap to fulfill its potential, which will also serve as a testbed for TAAP-based guided atom interferometry \cite{Pandey2019HypersonicRings, Navez2016}.
The non-linear nature of the magnetic field dependence on both DC and RF suggest the existence of a \textit{magic} set of parameters for the bi-chromatic shell trap clock transition, where the dependence on the strength of the RF-fields vanish linearly.

\vspace{2cm}
\section{Acknowledgements}

We gratefully acknowledge very fruitful discussions with Thomas Fernholz, Barry Garraway and German Sinuco. This work is supported by the project ``HELLAS-CH'' (MIS 5002735) which is
implemented under the  ``Action for Strengthening Research and Innovation
Infrastructures'', funded by the Operational Programme ``Competitiveness,
Entrepreneurship and Innovation'' (NSRF 2014-2020) and co-financed by
Greece and the European Union (European Regional Development Fund). SP acknowledge financial support from the Greek Foundation for Research and
Innovation (ELIDEK) in the framework of project, \emph{Guided Matter-Wave
Interferometry} under grant agreement number 4823 and General Secretariat
for Research and Technology (GSRT). GV received funding from the European Union’s Horizon 2020 research and innovation programme under the Marie Skłodowska-Curie Grant Agreement No  750017.
The author(s) would like to acknowledge the contribution of the COST Action CA16221.

\section{Contributions}

WK conceived the experimental ideas. 
HM, SP, GV carried out the experiments and performed the data analysis. 
HM, GV, WK developed the theory. 
All authors contributed to the result discussion and manuscript writing.

\appendix
\section{Radio-Frequency set-up for the bi-chromatic shell}%
\label{app:rfgeneration}

\begin{figure}[htb]
  \centering
  \begin{minipage}[c]{0.42\textwidth}
    \includegraphics[width=\textwidth]{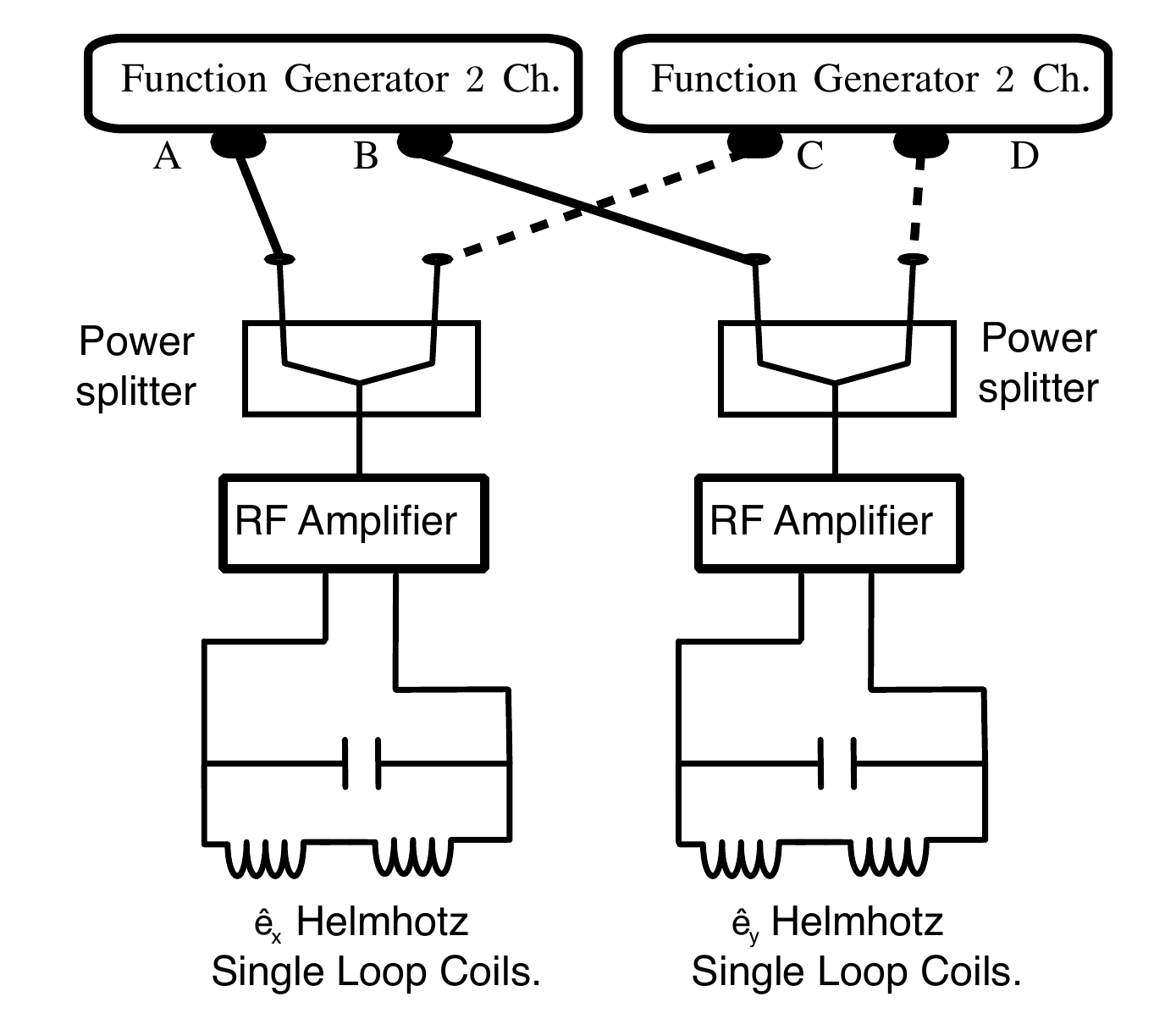}
  \end{minipage}
  \begin{minipage}[c]{0.38\textwidth}
    \centering
\begin{tabular}{ c | c }
Output & Waveform \\ \hline                        
A & $B_{\text{x1}}\cos{\left(\omega_{1}t\right)}$ \\
B & $-B_{\text{y1}}\sin{\left(\omega_{1}t+\phi_{1}\right)}$ \\
C & $B_{x2}\cos{\left(\omega_{2}t\right)}$ \\
D & $B_{y2}\sin{\left(\omega_{2}t+\phi_{2}\right)}$ \\
\end{tabular}
\end{minipage}
 \caption{Sketch of the RF generation set-up. Two commercial function generators feed two power splitters where the $\unitx$ and $\unity$ components of $B_{1}$ and $B_{\text{y}}$ are added and then fed, after amplification, to two pairs of single-loop Helmholtz coils oriented along the relevant axes, and tuned via resonators close to $2$\,MHz.}
\label{fig:rfgeneration}
\end{figure}

The cell in the experiment is surrounded by three pairs of Helmholtz coils in the $\unitx$, $\unity$ and $\unitz$ directions.
These have been tuned with LC circuits to a resonance with $Q\approx20$ at $\nu\approx{2.2}$\,MHz.
The desired RF-field consists of two circularly polarized components, at two different frequencies, with opposite handedness, as in Eq.\,\ref{eq:bifield}.
We feed each of the horizontal coils with the corresponding two-frequency component input, added from two function generators in separate power splitters.
This is depicted in Fig.\,\ref{fig:rfgeneration}.
The frequency generation sources are a Rigol DG4162 for the parameters with subindex 1 and a a Rigol DG1062 for the parameters with subindex 2.
The power splitters are a Minicircuits ZFRSC-42-S+ and ZFRSC-123-S+ and the amplifiers Minicircuits LZY-22+.
The microwave source used to drive the hyperfine transitions is a Rhodes and Swartz SMB 100A.
The microwave dipole antenna is home-made and tuned to the hyperfine line at $6.834$\,GHz.

\section{Trap loading}
\label{app:traploading}

\begin{figure}
\begin{center}
\resizebox{0.8\textwidth}{!}{%
  \includegraphics{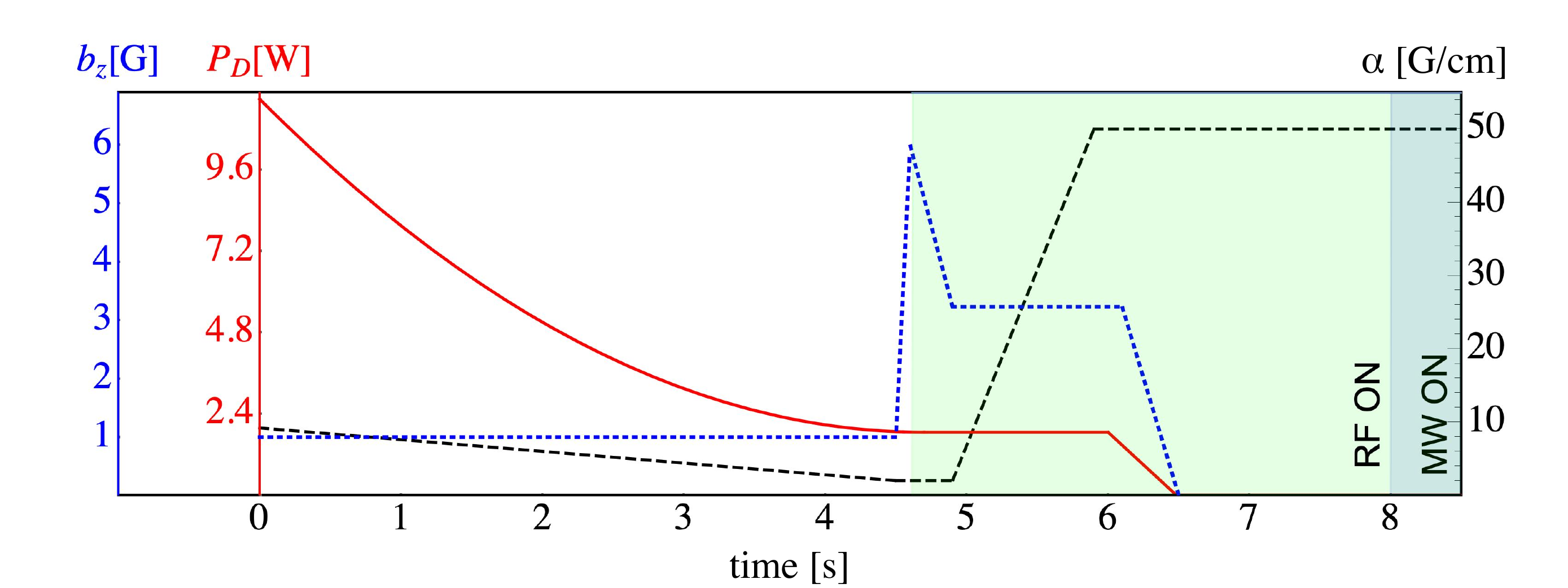}}
\caption{
Experimental sequence to load a shell
trap, with a focus on three key parameters: the vertical bias field $B_{\text{z}}$, the dipole laser 
power in the trap $P_{\text{D}}$, and the quadrupole field gradient $\alpha$. The light-green shadowed area indicates the duration of the RF-field radiation. The dark-turquoise shadowed area indicates the time when a MW pulse is switched on. Notice that the time duration of the latter is exaggerated for the readers' convenience. This figure illustrates the variation of these parameters from the onset of evaporation (where the solid red curve of $P_{\text{D}}$ begins its turn downwards) until right after a MW pulse is applied.
Following this sequence, the atom cloud is detected via absorption imaging. At $t\approx 4.5$\,s the RF is switched on and the cloud is dressed with a sweep of the bias field. At this point, the quadrupole is very weak, and atoms are still trapped in the crossed dipole. Two slow consecutive ramps, first of $\alpha$ to a high gradient field, then of $P_{\text{D}}$ to zero, result in the atoms trapped in either a mono-chromatic or bi-chromatic shell trap (depending on the whether the RF-field follows Eq.~\ref{eq:monochrombfield} or  Eq.~\ref{eq:bifield}, respectively).
}
\label{fig:expsequence}
\end{center}
\end{figure}

In a typical experiment (see Fig.~\ref{fig:expsequence}), we routinely load the RF-dressed shell trap from a crossed-beam optical dipole trap , that delivers up to $2\cdot10^{5}$ condensed atoms in $|1,-1\rangle$. To enter the dressed potential we first switch on a very weak quadrupole $\alpha<15$\,G/cm, which is displaced with a bias field - of magnitude $B_{z1}$- in the $\unitz$ direction. Afterwards we switch on $B_\text{RF}$  (RF-generation is described in \ref{app:rfgeneration}) and then slowly ramp down the bias field from $B_{z1}\rightarrow B_{z2}$ ($\Delta t_{B_{z1}\rightarrow B_{z2}}=400$\,ms) until the crossed dipole trap and the shell trap are overlapped. $B_{z1},B_{z2}$ are chosen according to $\hbar \omega_{1}\approx \mu_{\text{B}}|g_{1}| B_{z2} = \mu_{\text{B}}|g_{1}| B_{z1}/2$. In the final step, we ramp down the dipole trap laser power ($\Delta t_{\textrm{transfer}}=200$\,ms) and switch it off. We can load clouds without noticeable heating nor atom number loss. We have measured up to a 7\,s long lifetime for a condensate and up to 80s lifetime (the vacuum lifetime is 2\,min) for a $T<200$\,nK thermal cloud in the bi-chromatic shell trap. Trapping frequency measurements in both the mono-chromatic shell and the bi-chromatic shell, with similar experimental parameters, agree with Eqs.(\ref{eq:shelltrapfreqrad},\ref{eq:shelltrapfreqaxial}), being typically in the ranges: $\omega_\text{z}/2\pi$ from $50$\,Hz to $300$\,Hz and $\omega_{\rho}/2\pi$ from $8$\,Hz to $13$\,Hz.

\section{Multi-photon spectrum}
\label{app:multiphoton}

The multi-photon spectrum can be explained by imperfections in the RF-polarization. We will derive an approximate expression to describe the resulting eigenenergies modulation due to cross-coupling from the RF-dressing field of the \textit{opposite} manifold. In order to account for imperfections in the RF-fields polarization we will find the circular projections of $B_\text{1}$ ($B_\text{2}$) in $\sigma_{-}$ ($\sigma_{+}$) into the local basis of the quadrupole at the center of the traps, where it is aligned with $\unitz$. We will assume that the phase at the RF generation coils is $s \pi/2$, but that the RF-field amplitudes in the two coils are not the same. Then, for a generic case, we have:

\begin{equation}
B_{\textrm{RF}}=B_{\text{x}} \cos{(\omega t)} \unitx + B_{\text{y}} \sin{(\omega t)}\unity
\end{equation}

\noindent Now we can use that $B=(B_{\text{x}}+B_{\text{y}})/2$ and $\Delta B=(B_{\text{x}}-B_{\text{y}})/2$ to write:

\begin{equation}
B_{\textrm{RF}}=(B-\Delta_{B}) \cos{(\omega t)} \unitx + (B+\Delta_{B})  \sin{(\omega t)}\unity
\end{equation}

\noindent where one can readily see a counter-rotating component of amplitude $\Delta B$. This will result in a total field with crossed components $\Omega_{1,2}=\frac{|g_{F}|\mu_{\text{B}}}{2 \hbar}\Delta B_{B1}$ and $\Omega_{2,1}=\frac{|g_{F}|\mu_{\text{B}}}{2 \hbar}\Delta B_{B2}$, where $\Delta B_{B1}=(B_{\text{x1}}-B_{\text{y1}})/2$ and $\Delta B_{B2}=(B_{x2}-B_{y2})/2$. We can now write the full magnetic field as:

\begin{equation}
\begin{split}
\frac{g_{F}\mu_{\text{B}}}{2\hbar} \mathbf{B}_\text{RF}(t)=
((\Omega_{1}-\Omega_{1,2})\cos{(\omega_{1}t)}+(\Omega_{2}-\Omega_{2,1})\cos{(\omega_{2}t)})\unitx\\+
(-(\Omega_{1}+\Omega_{1,2})\sin{(\omega_{1}t)}+(\Omega_{2}+\Omega_{2,1})\sin{(\omega_{2}t)})\unity
\end{split}
\end{equation}

\noindent where $\Omega_{j}=\frac{g_{F}\mu_{\text{B}}}{2\hbar}B_j$. Notice that $\Omega_{1,2},\Omega_{2,1}$ are the crossed contributions from $B_\text{rf}^1$ into $F=2$ and from $B_\text{rf}^2$ into $F=1$. The static part of the Hamiltonian in Eq.~\ref{eq:H_uncoupled} is:

\begin{figure}
\begin{center}
\resizebox{0.5\textwidth}{!}{\includegraphics{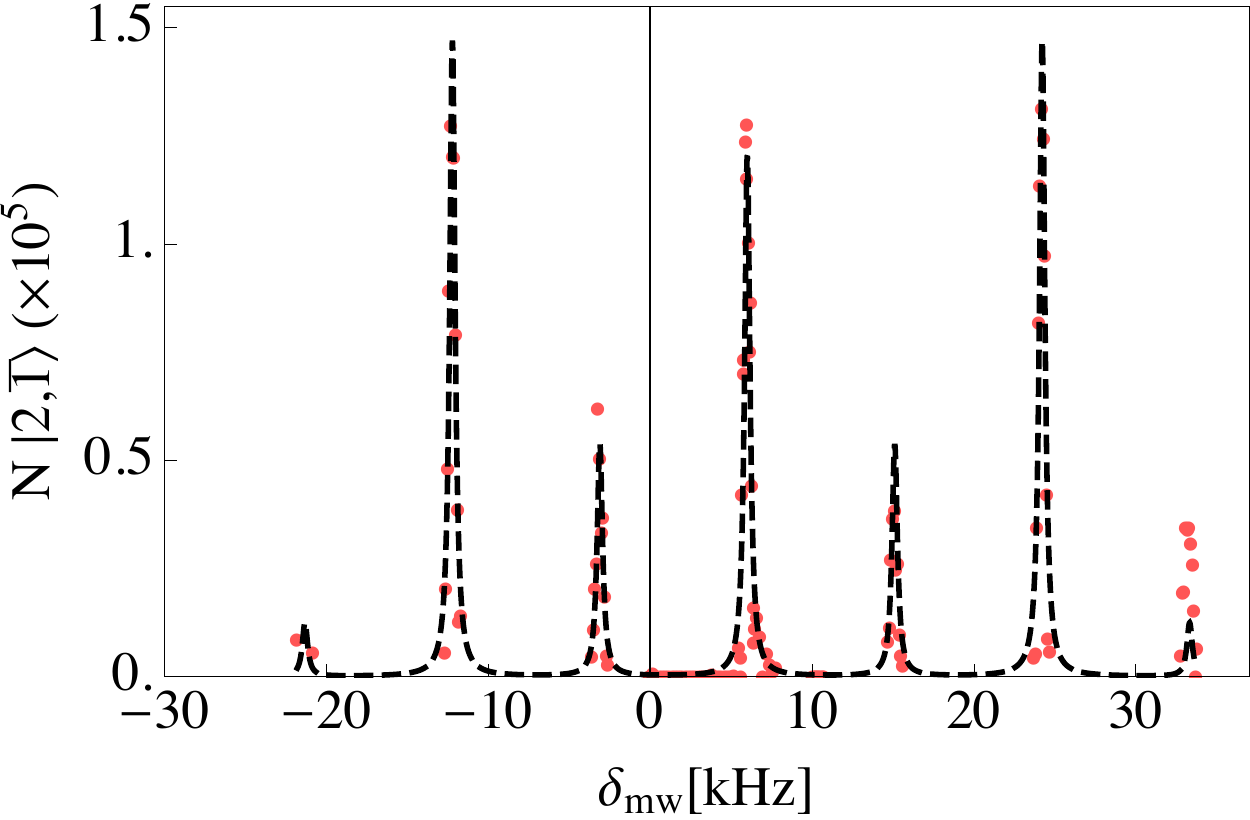}}
\caption{An example of a multi-photon spectrum. Red dots indicate experimental measurements and the Black Dashed line is a fit to the data employing the result in Eq.~\ref{eq:spectrum}}
\label{fig:multifreqspec}
\end{center}
\end{figure}

\begin{equation}
H_{0}=H_{\text{hfs}}+H_{\text{DC}}=\sum_{F=1,2} \frac{A_{\text{hfs}}}{2}(F(F+1)-9/2)\hat{P}^{F}+\frac{\mu_{\text{B}}}{\hbar}\sum_{F=1,2} g_{F}\fz^{F} B_{\text{DC}}
\label{eq:staticpart}
\end{equation}

\noindent where $P^{1}$, $P^{2}$ ($\fz^{1}$,$\fz^{2}$) are projectors (operators) into the respective two subspaces with the standard definitions \cite{Perrin2017}. Here we have chosen a $B_{\text{DC}}$ that would correspond to the quadrupole field at $z_{0}$. The time dependent part is $H_\text{RF}=\mathbf{F}\boldsymbol{B}_\text{RF}(t)$. In the absence of a strong coupling between the hyperfine manifolds, and assuming that the atoms in the sub-states $|F,\mFD\rangle$ follow adiabatically Eq.~\ref{eq:dressedPotentiala}, we can calculate the time evolution in the state-dependent rotating frames resulting from the application of:

\begin{equation}
\zRot (t)=\prod_{F} \exp{\left(\frac{-i \omega_{F} t \fz^{F}}{s\hbar}\right)}
\label{eq:rotation}
\end{equation}

\noindent where we have assumed, for simplicity, that $g_{1}=-g_{2}$. We calculate the time evolution $\hat{H}=\zRot^{\dagger}(H_{0}+H_\text{RF})\zRot-i \hbar \zRot^{\dagger} \delta_{t} \zRot$, after which we define $\Delta =\omega_{1}-\omega_{2}$, similarly to \cite{GarridoAlzar2006}. The resulting Hamiltonian is:

\begin{equation}
\begin{aligned}
H=H_{\text{hfs}}+
\hbar^{-1}\mF s\left(|g_{F}| \mu_{\text{B}} B_{\text{DC}}- s\hbar \omega_{F} \right)\fz^{F}+\\
\sum_{F,F'} \left[\delta_{F,F'}\Omega_{F} \fx^{F}+
(1-\delta_{F,F'})\Omega_{F,F'} \left[\fx^{F'}\cos{(\Delta t)}-\fy^{F'}\sin{(\Delta t)}\right]\right]
\end{aligned}
\end{equation}

\noindent where the sum is over $\left(F,F'\right)=1,2$ and $\delta_{F,F'}=0,\delta_{F,F}=1$. Close to resonance, $\mFD$ remain \textit{good} quantum numbers and we can find the eigenenergies:

\begin{equation}
E_{F,\bar{m}}\left(t\right)=s\left(I+\frac{1}{2}\right)\frac{\hbar \omega_\text{hfs}}{2}+ |\mFD| \hbar \sqrt{\delta_{F}^{2} + \Omega_{F,\mFD}^{2}\left(t\right)}
\label{eq:modeigenstates}
\end{equation}

\noindent with $s=1$ ($s=-1$) for $F=2$ ($F=1$) and $I=3/2$ and where $\delta_{F}=\mu_B |g_F| B_{\text{DC}} - \hbar \omega_\text{F}$, and:

\begin{equation}
\Omega_{F,\mFD}^2(t)=\Omega_{F}^2+\Omega_{F,F'}^2+2\Omega_{F}\Omega_{F,F'}\cos{(\Delta t)}
\label{eq:modfields}
\end{equation}

\noindent For small $\Omega_{2,1}$, $\Omega_{1,2}$ the time averaged potential resulting from Eq.~\ref{eq:modeigenstates} is well approximated by Eq.~\ref{eq:dressedPotentiala}. Notice however that we did not account for the spatial dependence of the quadrupole field, since we have assumed a homogeneous field aligned with $\unitz$, which is only a fair approximation for small clouds at the bottom of the trap. It is nonetheless illustrative enough to now develop this argument further by considering that any pair of $|1,\mFD \rangle ,|2,\bar{m}'_{F}\rangle$ can be treated as a two level system  coupled by a weak microwave field. Particularly, in this two-level atom picture, for small $\Omega_{1,2},\Omega_{2,1}$, the difference in energy between $|1,-1\rangle$ and $|2,1\rangle$ the two states will be modulated by:

\begin{equation}
\Delta \Omega (t)=\Omega_{2}-\Omega_{1}+(\Omega_{2,1}-\Omega_{1,2})\cos{(\Delta t)}
\end{equation}

\noindent and now we can write the reduced matrix  in a two-dimensional basis with  $|2,1\rangle \rightarrow |\uparrow\rangle$ and $|1,-1\rangle \rightarrow |\downarrow\rangle$:

\begin{equation}
H_{m}/\hbar= \left(\frac{\whf}{2}+\frac{\Delta \Omega (t)}{2} \right)\hat{\sigma}_{z}+ \Omega_{\text{MW}} \cos{(\wMW t)} \hat{\sigma}_{\text{x}}
\end{equation}

\noindent where we have included a weak microwave link between the two states. This system is described in detail in ~\cite{Silveri2017}, from where we retrieve the general solution for the spectral response:

\begin{equation}
\label{eq:spectrum}
P_{\omega} \left(\omega\right)=\sum_{j=-\infty}^{\infty}\frac{\Gamma|\Delta_\text{MW} J_{j}\left(\xi\right)|^{2}}{\left(\delta+j \Delta \omega\right)^{2}+\Gamma^{2}}
\end{equation}

\noindent where $\Gamma$ is the line-width of each transition, determined by all the involved decoherence processes, and $\delta=\omega_\text{MW}-\omega_{\text{RF}}$, $\Delta \omega =\omega_{1}-\omega_{2}$.
This expression corresponds to an infinite number of transitions centered at $\Omega_{\text{RF}}$ and spaced by $\Delta \omega$ with relative amplitudes given by the family of Bessel functions of the first kind with argument $\xi=\Delta\Omega_{F,F+s}/\Delta \omega$, where $\Delta\Omega_{F,F+s}=|\Omega_{1,2}-\Omega_{2,1}|$ is the amplitude of the energy splitting modulation. 
In the bi-chromatic shell spectrum, all 7 transitions corresponding to $n=-3...,0,...3$ will present this multi-photon resonance. Fig.~\ref{fig:multifreqspec} shows an example of a multi-photon spectrum measured in the bi-chromatic shell. We can fit the data, depicted in Red dots, to Eq.~\ref{eq:spectrum} to find an modulation amplitude of the two levels eigenenergies of $|\Omega_{2,1}-\Omega_{1,2}|/2\pi=118$\,kHz and $\Delta/2\pi=9098$\,Hz for $\Omega_{1}/2\pi\approx\Omega_{2}/2\pi=123$\,kHz and $\omega_{1}/2\pi=2.294336$\,MHz and $\omega_{2}/2\pi=2.285238$\,MHz.

\section*{References}
 \bibliographystyle{unsrt}

\begin{thebibliography}{10}

\bibitem{Rosi2014N}
G.~Rosi, F.~Sorrentino, L.~Cacciapuoti, M.~Prevedelli, and G.~M. Tino.
\newblock Precision measurement of the newtonian gravitational constant using
  cold atoms.
\newblock {\em Nature}, 510(518--521):--, 06 2014.

\bibitem{Hamilton2015s}
P~Hamilton, M~Jaffe, P~Haslinger, Q~Simmons, H~M{\"{u}}ller, and J~Khoury.
\newblock {Atom-interferometry constraints on dark energy.}
\newblock {\em Science (New York, N.Y.)}, 349(6250):849--51, 8 2015.

\bibitem{Margalit2015S}
Yair Margalit, Zhifan Zhou, Shimon Machluf, Daniel Rohrlich, Yonathan Japha,
  and Ron Folman.
\newblock A self-interfering clock as a ``which path'' witness.
\newblock {\em Science}, 349(6253):1205--1208, 2015.

\bibitem{Manning2015NP}
A.~G. Manning, R.~I. Khakimov, R.~G. Dall, and A.~G. Truscott.
\newblock Wheeler's delayed-choice gedanken experiment with a single atom.
\newblock {\em Nature Physics}, 11(7):539--U142, JUL 2015.

\bibitem{Jaffe2017NP}
Matt Jaffe, Philipp Haslinger, Victoria Xu, Paul Hamilton, Amol Upadhye,
  Benjamin Elder, Justin Khoury, and Holger Mueller.
\newblock Testing sub-gravitational forces on atoms from a miniature in-vacuum
  source mass.
\newblock {\em Nature Physics}, 13(10):938+, OCT 2017.

\bibitem{Geraci2016PRL}
Andrew~A. Geraci and Andrei Derevianko.
\newblock Sensitivity of atom interferometry to ultralight scalar field dark
  matter.
\newblock {\em Physical Review Letters}, 117(26), DEC 20 2016.

\bibitem{Zhou2015PRL}
Lin Zhou, Shitong Long, Biao Tang, Xi~Chen, Fen Gao, Wencui Peng, Weitao Duan,
  Jiaqi Zhong, Zongyuan Xiong, Jin Wang, Yuanzhong Zhang, and Mingsheng Zhan.
\newblock Test of equivalence principle at $1{0}^{\ensuremath{-}8}$ level by a
  dual-species double-diffraction raman atom interferometer.
\newblock {\em Phys. Rev. Lett.}, 115:013004, Jul 2015.

\bibitem{Aguilera2014GAQG}
DN~Aguilera, H~Ahlers, Baptiste Battelier, Ahmad Bawamia, Andrea Bertoldi,
  R~Bondarescu, K~Bongs, Philippe Bouyer, C~Braxmaier, L~Cacciapuoti, et~al.
\newblock Ste-quest—test of the universality of free fall using cold atom
  interferometry.
\newblock {\em Classical and Quantum Gravity}, 31(11):115010, 2014.

\bibitem{Bidel2018NC}
Y.~Bidel, N.~Zahzam, C.~Blanchard, A.~Bonnin, M.~Cadoret, A.~Bresson,
  D.~Rouxel, and M.~F. Lequentrec-Lalancette.
\newblock Absolute marine gravimetry with matter-wave interferometry.
\newblock {\em Nature Communications}, 9(1), Feb 2018.

\bibitem{Becker2018N}
Dennis Becker, Maike~D. Lachmann, Stephan~T. Seidel, Holger Ahlers, Aline~N.
  Dinkelaker, Jens Grosse, Ortwin Hellmig, Hauke M{\"u}ntinga, Vladimir
  Schkolnik, Thijs Wendrich, and et~al.
\newblock Space-borne bose--einstein condensation for precision interferometry.
\newblock {\em Nature}, 562(7727):391--395, Oct 2018.

\bibitem{Zoest2010S}
T.~van Zoest, N.~Gaaloul, Y.~Singh, H.~Ahlers, W.~Herr, S.~T. Seidel,
  W.~Ertmer, E.~Rasel, M.~Eckart, E.~Kajari, S.~Arnold, G.~Nandi, W.~P.
  Schleich, R.~Walser, A.~Vogel, K.~Sengstock, K.~Bongs, W.~Lewoczko-Adamczyk,
  M.~Schiemangk, T.~Schuldt, A.~Peters, T.~K{\"o}nemann, H.~M{\"u}ntinga,
  C.~L{\"a}mmerzahl, H.~Dittus, T.~Steinmetz, T.~W. H{\"a}nsch, and J.~Reichel.
\newblock Bose-einstein condensation in microgravity.
\newblock {\em Science}, 328(5985):1540--1543, 06 2010.

\bibitem{Barrett2016NC}
Brynle Barrett, Laura Antoni-Micollier, Laure Chichet, Baptiste Battelier,
  Thomas L{\'e}v{\`e}que, Arnaud Landragin, and Philippe Bouyer.
\newblock Dual matter-wave inertial sensors in weightlessness.
\newblock {\em Nature Communications}, 7:13786, Dec 2016.

\bibitem{Weidner2018}
C~A Weidner and Dana~Z Anderson.
\newblock {Experimental Demonstration of Shaken-Lattice Interferometry}.
\newblock 2018.

\bibitem{Zhang2016}
Xian Zhang, Ruben~Pablo del Aguila, Tommaso Mazzoni, Nicola Poli, and
  Guglielmo~M. Tino.
\newblock {Trapped-atom interferometer with ultracold Sr atoms}.
\newblock {\em Physical Review A}, 94(4):043608, 10 2016.

\bibitem{Andia2013PRA}
Manuel Andia, Rapha{\"e}l Jannin, Fran{\c c}ois Nez, Fran{\c c}ois Biraben,
  Sa{\"\i}da Guellati-Kh{\'e}lifa, and Pierre Clad{\'e}.
\newblock Compact atomic gravimeter based on a pulsed and accelerated optical
  lattice.
\newblock {\em Physical Review A}, 88(3), Sep 2013.

\bibitem{Bohi2010APL}
P.~Bohi, M.F. Riedel, T.W. Hansch, and P.~Treutlein.
\newblock {Imaging of microwave fields using ultracold atoms}.
\newblock {\em Applied Physics Letters}, 97(5):051101--051101, 2010.

\bibitem{Gierling2011NN}
M.~Gierling, P.~Schneeweiss, G.~Visanescu, P.~Federsel, M.~H{\"a}ffner, D.~P.
  Kern, T.~E. Judd, A.~G{\"u}nther, and J.~Fort{\'a}gh.
\newblock Cold-atom scanning probe microscopy.
\newblock {\em Nature Nanotechnology}, 6(7):446--451, May 2011.

\bibitem{Steck2001}
Daniel~A Steck.
\newblock {Rubidium 87 D Line Data (http://steck.us/alkalidata)}, 2001.

\bibitem{Zobay2001}
O.~Zobay and B.~M. Garraway.
\newblock {Two-dimensional atom trapping in field-induced adiabatic
  potentials}.
\newblock {\em Physical Review Letters}, 86(7):1195--1198, 2001.

\bibitem{Garraway2015}
Barry Garraway and Helene Perrin.
\newblock {Recent developments in trapping and manipulation of atoms with
  adiabatic potentials}.
\newblock {\em J. Phys. B: At. Mol. Phys. on}, 49(17):172001, 2015.

\bibitem{Lesanovsky2006}
I.~Lesanovsky, S.~Hofferberth, J.~Schmiedmayer, and P.~Schmelcher.
\newblock {Manipulation of ultracold atoms in dressed adiabatic radio-frequency
  potentials}.
\newblock {\em Physical Review A}, 74(1):033619, 2006.

\bibitem{Merloti2013}
K~Merloti, R~Dubessy, L~Longchambon, A~Perrin, P-E Pottie, V~Lorent, and
  H~Perrin.
\newblock {A two-dimensional quantum gas in a magnetic trap}.
\newblock {\em New Journal of Physics}, 15(3):033007, 3 2013.

\bibitem{Navez2016}
P.~Navez, S.~Pandey, H.~Mas, K.~Poulios, T.~Fernholz, and W.~Von Klitzing.
\newblock {Matter-wave interferometers using TAAP rings}.
\newblock {\em New Journal of Physics}, 18(7):075014, 2016.

\bibitem{Haroche1970}
S.~Haroche, C.~Cohen-Tannoudji, C.~Audoin, and J.~P. Schermann.
\newblock {Modified Zeeman Hyperfine Spectra Observed in H 1 and Rb 87 Ground
  States Interacting with a Nonresonant rf Field}.
\newblock {\em Physical Review Letters}, 24(16):861--864, 4 1970.

\bibitem{Nottingham}
G.~A. Sinuco-Leon, B.~M. Garraway, H.~Mas, S.~Pandey, G.~Vasilakis, V.~Bolpasi,
  W.~von Klitzing, B.~Foxon, S.~Jammi, K.~Poulios, and T.~Fernholz.
\newblock {Microwave spectroscopy of radio-frequency dressed
  {\$}{\^{}}{\{}87{\}}{\$}Rb}.
\newblock {\em ArXiv ID 1904.12073}, 4 2019.

\bibitem{Luksch2018}
Kathrin Luksch, Elliot Bentine, Adam~J. Barker, Shinichi Sunami, Tiffany~L.
  Harte, Ben Yuen, and Christopher~J. Foot.
\newblock {Probing multiple-frequency atom-photon interactions with ultracold
  atoms}.
\newblock 12 2018.

\bibitem{Silveri2017}
M.~P. Silveri, J.~A. Tuorila, E.~V. Thuneberg, and G.~S. Paraoanu.
\newblock {Quantum systems under frequency modulation}.
\newblock {\em Reports on Progress in Physics}, 80(5):056002, 5 2017.

\bibitem{Oliver2005}
William~D. Oliver, Yang Yu, Janice~C. Lee, Karl~K. Berggren, Leonid~S. Levitov,
  and Terry~P. Orlando.
\newblock {Mach-Zehnder Interferometry in a Strongly Driven Superconducting
  Qubit}.
\newblock {\em Science}, 310(5754):1653--1657, 2005.

\bibitem{GarridoAlzar2006}
Carlos~L. Garrido~Alzar, Helene Perrin, Barry~M. Garraway, and Vincent Lorent.
\newblock {Evaporative cooling in a radio-frequency trap}.
\newblock {\em Physical Review A}, 74(5):053413, 2006.

\bibitem{Pandey2019HypersonicRings}
Saurabh Pandey, Hector Mas, Giannis Drougakis, Premjith Thekkeppatt, Vasiliki
  Bolpasi, Georgios Vasilakis, Konstantinos Poulios, and Wolf von Klitzing.
\newblock {Hypersonic Bose--Einstein condensates in accelerator rings}.
\newblock {\em Nature}, 570:205--209, 2019.

\bibitem{Perrin2017}
H\'el\`ene Perrin and Barry~M. Garraway.
\newblock {Trapping Atoms With Radio Frequency Adiabatic Potentials}.
\newblock {\em Advances In Atomic, Molecular, and Optical Physics},
  66:181--262, 2017.

\end{thebibliography}

\end{document}